\documentclass[twocolumn,
               superscriptaddress,
               floatfix,
               longbibliography, 
               showkeys,apl]{revtex4-1}

\usepackage{graphicx} 
\usepackage{bm}	
\usepackage{color}                     
\usepackage{xcolor}
\usepackage{epsfig}
\usepackage{amsmath} 
\usepackage{amssymb} 
\usepackage{longtable} 
\usepackage{float}
\usepackage{mathtools}
\usepackage{xparse}
\usepackage{hyperref}
\usepackage{times}
\usepackage[normalem]{ulem}
\usepackage[ruled]{algorithm}
\usepackage{algpseudocode}
\usepackage{enumitem}

\usepackage{sidecap}
\sidecaptionvpos{figure}{t}

\definecolor{darkgreen}{rgb}{0.13, 0.55, 0.13}
\def\new#1{ #1}

\newcommand{\orquestra}{\mbox{ORQUESTRA\textsuperscript{\textregistered}}\ }
\newcommand{\sufour}{SU$(4)$\ }
\newcommand{\params}{{\bm{\theta}}}

\newcommand{\braket}[2]{\left\langle#1|#2\right\rangle}

\begin{document}

\setlist[enumerate,1]{label=\arabic*, start=0}


\title{Synergy Between Quantum Circuits and Tensor Networks:\\ Short-cutting the Race to Practical Quantum Advantage}

\author{Manuel S. Rudolph}
\affiliation{Zapata Computing Canada Inc., 325 Front St W, Toronto, ON, M5V 2Y1}

\author{Jacob Miller}
\affiliation{Zapata Computing Canada Inc., 325 Front St W, Toronto, ON, M5V 2Y1}

\author{Danial Motlagh}
\affiliation{Zapata Computing Canada Inc., 325 Front St W, Toronto, ON, M5V 2Y1}

\author{Jing Chen}
\affiliation{Zapata Computing Inc., 100 Federal Street, Boston, MA 02110, USA}

\author{Atithi Acharya}
\affiliation{Zapata Computing Inc., 100 Federal Street, Boston, MA 02110, USA}
\affiliation{Rutgers University, 136 Frelinghuysen Rd, Piscataway, NJ 08854, USA}

\author{Alejandro Perdomo-Ortiz}
\email{alejandro@zapatacomputing.com}
\affiliation{Zapata Computing Canada Inc., 325 Front St W, Toronto, ON, M5V 2Y1}

\date{\today} 

\begin{abstract}

  While recent breakthroughs have proven the ability of noisy intermediate-scale quantum (NISQ) devices to achieve quantum advantage in classically-intractable sampling tasks, the use of these devices for solving more practically relevant computational problems remains a challenge. Proposals for attaining practical quantum advantage typically involve parametrized quantum circuits (PQCs), whose parameters can be optimized to find solutions to diverse problems throughout quantum simulation and machine learning.
  However, training PQCs for real-world problems remains a significant practical challenge, largely due to the phenomenon of barren plateaus in the optimization landscapes of randomly-initialized quantum circuits.
  In this work, we introduce a scalable procedure for harnessing classical computing resources to \new{provide pre-optimized} 
  initializations \new{for} PQCs, which we show significantly improves the trainability and performance of PQCs on a variety of problems. Given a specific optimization task, this method first utilizes tensor network (TN) simulations to identify a promising quantum state, which is then converted into gate parameters of a PQC by means of a high-performance decomposition procedure. We show that this \new{learned initialization} avoids barren plateaus, and effectively translates increases in classical resources to enhanced performance and speed in training quantum circuits. By demonstrating a means of boosting limited quantum resources using classical computers, our approach illustrates the promise of this synergy between quantum and quantum-inspired models in quantum computing, and opens up new avenues to harness the power of modern quantum hardware for realizing practical quantum advantage.
  
  \end{abstract}

\maketitle

\section{Introduction}\label{sec:intro}

\begin{figure*}
    \centering
    \includegraphics[width=0.78\linewidth]{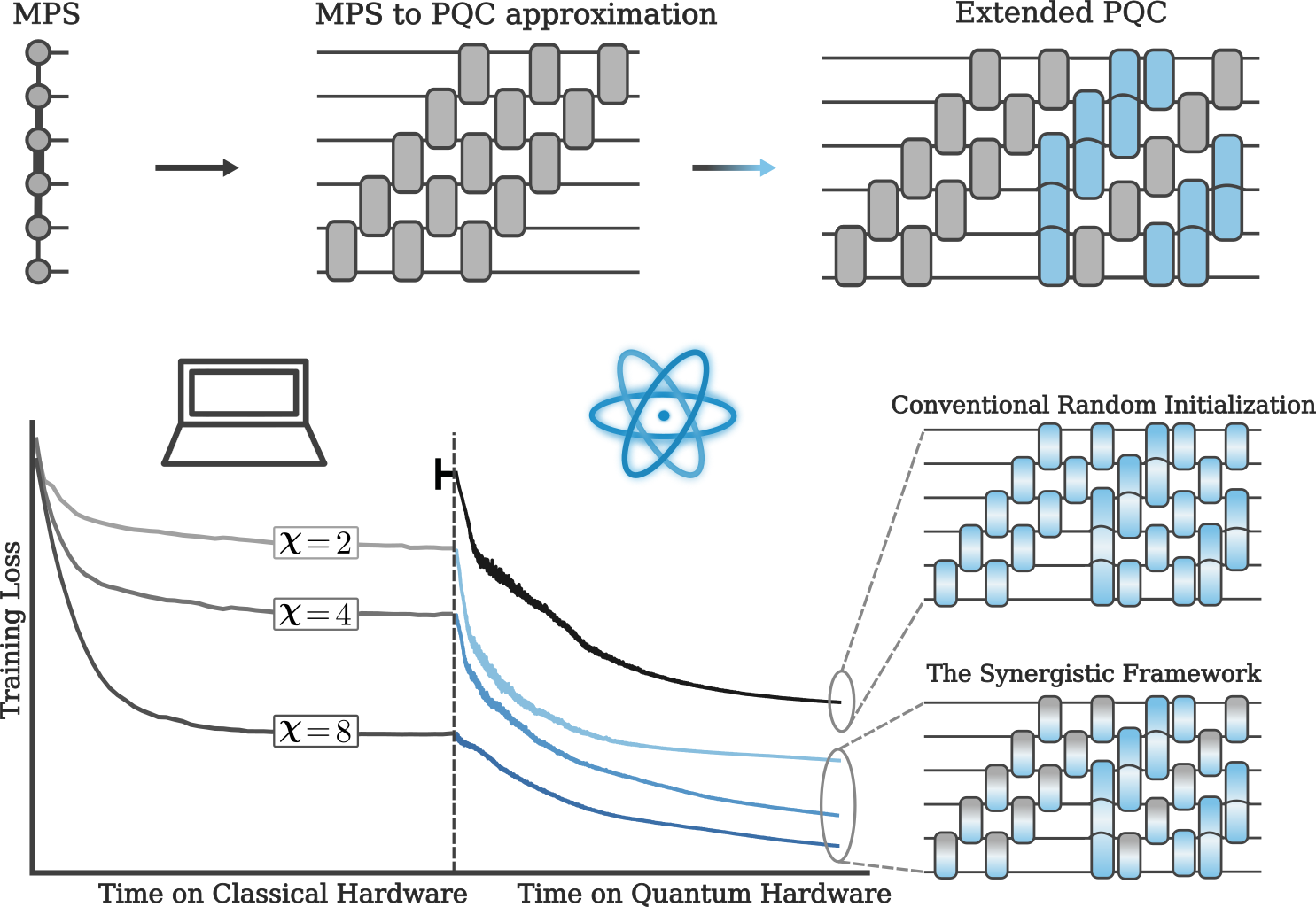}
    \caption{Schematic depiction of the synergistic training framework utilizing TNs and PQCs. Rather than starting with a random initialization of circuit parameters (black curve), which may suffer from problems such as barren plateaus and sub-optimal local minima, we instead train a matrix product state (MPS) model on a classical simulation of the problem at hand (left half of blue curves), whose performance is bounded by the limited entanglement available via its bond dimension $\chi$. This MPS wavefunction is then approximately transferred using a layer-efficient decomposition protocol that maps the MPS to linear layers of \sufour gates. To improve on the classical solution, the quantum circuit is extended with additional gates (blue gates, initialized as near-identity operations) that would have been unfeasible to simulate on classical hardware. 
    We show that quantum circuit models that leverage classically initialized circuit layers (gray \& blue shaded gates) exhibit drastically improved performance over quantum circuits that were fully optimized on quantum hardware (blue shaded gates) and are likely to run into common trainability issues.
    }
    \label{fig:carton_example}
\end{figure*}

In the coming years and decades, quantum computing resources will likely remain more expensive and less abundant than classical computing resources~\cite{Aaronson-2015, Preskill2018, national2019quantum}. 
Despite the intrinsic theoretical advantages of quantum computers, the widespread adoption of quantum technologies will ultimately depend on the benefits they can offer for solving problems of high practical interest using these limited resources. To this end, parametrized quantum circuits (PQCs)~\cite{benedetti2019parameterized,Cerezo2021_vqareview,Bharti2022_nisqreview} have been proposed as a promising formalism for leveraging near-term quantum devices for the solution of problems in quantum chemistry~\cite{Cao2019,mcardle2020quantum,lubasch2020variational}, materials science~\cite{bauer2020quantum}, and quantum machine learning~\cite{ biamonte2016quantum, PerdomoOrtiz2017, Farhi2018, Benedetti2019, 2019schuld,Havlicek2019, rudolph2020generation,Huang2022} applications which are difficult for classical algorithms.

However, several distinct challenges stand in the way of reaching practical advantage over classical methods using parametrized quantum algorithms, such as the existence of \textit{barren plateaus}~\cite{Mcclear2018Barren,cerezo2021costfunction,holmes2022expressivity,wang2021noiseinduced} and unfavorable guarantees for local minima~\cite{anschuetz2022beyond,anschuetz2021critical,arrasmith2021gorges} in the PQC optimization landscape. Such results typically apply to the setting of generic PQC ansätze and parameter initialization schemes, and much less is known about scenarios where the initial parameters of a PQC are adapted for a particular task. While this adaptation has proven useful in quantum chemistry, where circuits for computing molecular ground states have been shown to reach higher-precision results using initializations based on mean-field Hartree-Fock or more sophisticated coupled-cluster-based solutions (e.g., see Refs.~\cite{romero2019strategies,wecker2015progress,kottmann2022optimized,kottmann2022molecular}), task-specific initializations have seen much less use in other areas, such as \textit{quantum machine learning} (QML).

Another difficulty for demonstrating an advantage over classical algorithms using PQCs is the increasing sophistication of classical simulation algorithms based on \textit{tensor networks} (TNs), whose classically parametrized models can efficiently describe PQCs whose intermediate states have limited entanglement. The ability of TNs to be deployed on powerful classical hardware accelerators, such as graphical and tensor processing units (GPUs and TPUs), raises the bar for quantum hardware to overcome. This situation has led to a ``zero-sum game'' perspective on improvements in quantum vs. classical technologies, where advances in one domain are frequently viewed as placing additional burdens on practitioners of the other to attain relative advantages (see Ref.~\cite{cho2022ordinary} for a representative example).

As depicted in Fig.~\ref{fig:carton_example}, we propose a synergistic framework for boosting the performance and trainability of PQCs using a \new{pre-optimized} initialization strategy built on scalable TN algorithms, which leverages the complementary strengths of both technologies. This method uses TNs to first find a promising quantum state for the parametrized quantum algorithm at hand, then converts this TN state to the parameters of a PQC, where further optimization can be carried out on quantum hardware. We employ a circuit layer-efficient decomposition protocol~\cite{rudolph2022MPSdecomposition} for matrix product states (MPS), whose high-\new{fidelity} conversion of MPS to various PQC architectures allows leveraging high-quality MPS solutions. The resulting quantum circuits can be extended with classically infeasible gates which enable better performance relative to the MPS, as well as purely quantum-optimized circuits. We empirically verify these performance improvements in various problems from generative modeling and Hamiltonian ground state search, finding that our method successfully converts deep quantum circuits from being practically untrainable to reliably converging to high-quality solutions. We further give evidence for the scalability of our synergistic framework by probing the gradient variances, i.e., the ``barrenness'', of PQCs with up to \new{100} qubits, finding gradient variances \new{and magnitudes} to remain stable with increasing number of qubits and circuit depth. By ensuring that PQCs are set up for success using the best solution available with today's abundant classical computing resources, we believe that \new{our methods might finally unlock the true potential of} parametrized quantum algorithms as effective methods for solving problems of deep practical interest.

\section{Related Work}


Compared to previous works, our method is broadly similar to the proposal of Ref.~\cite{Huggins_2019} to use classically trained TN models for initializing PQCs, which was predicted to yield benefits in performance and trainability within general machine learning tasks. Our findings can thus be seen as both a concrete realization of this general proposal, applicable to a diverse range of circuit architectures and learning tasks, as well as a robust experimental verification of the benefits anticipated there. Closer to our work is the pretraining method of Ref.~\cite{dborin2022pretraining}, where trained MPS with bond dimension $\chi=2$ were exactly decomposed into a staircase of two-qubit gates, which were then used to initialize quantum circuits for machine learning tasks. While this method was shown to improve the performance and trainability of PQC models, the restriction to $\chi=2$ MPS placed a limit on the extent of classical resources which could be used to improve quantum models. By contrast, our synergistic optimization framework can be scaled up to utilize arbitrarily large classical and quantum resources, a difference that we show gives continued returns in practice.

\new{While the method we develop utilizes the specific circuit decomposition procedure of Ref.~\cite{rudolph2022MPSdecomposition}, any other scalable MPS to PQC decomposition can be used in its place, so long as the following criteria are met: (a) It must accept as input MPS of arbitrarily large bond dimensions, (b) It must output a circuit of any desired depth formed from one- and two-qubit gates, and (c) It must converge to the original MPS state vector at a reasonable rate as the circuit depth increases. All of these criteria must be satisfied for the method to deliver the benefits seen here, with criterion (a) needed to use increasing classical resources (Ref.~\cite{dborin2022pretraining} is limited here), criterion (b) needed to use increasing quantum resources within real-world quantum computers (the methods of Refs.~\cite{schon2005sequential, cramer2010efficient, gundlapalli2022deterministic} output single-layer circuits of $k$-qubit gates, with $k$ unbounded), and criterion (c) needed to avoid fidelity plateaus which hinder the conversion of high-quality MPS into high-quality PQC (Ref.~\cite{ran2020encoding} exhibits such fidelity plateaus~\cite{rudolph2022MPSdecomposition}). Besides Ref.~\cite{rudolph2022MPSdecomposition}, also the decomposition algorithms in Refs.~\cite{zhou2021automatically, dov2022approxencoding} satisfy all of these criteria, and are therefore promising candidates to be employed within this synergistic optimization framework.}


\section{Methods}\label{sec:methods}
\subsection{Parametrized Quantum Circuits}\label{ssec:methods_PQCs}
\textit{Parametrized quantum circuits} (PQCs), and in particular the so-called \textit{variational quantum algorithms} (VQAs),  are the centerpiece of a family of quantum algorithms that aim to solve practical problems on near-term quantum devices with a limited number of qubits and noisy operations. The parameters $\params$ of PQCs are usually optimized (or ``trained'', in the context of learning from data) according to a loss function 
\begin{equation}\label{eq:loss_function}
    \mathcal{L}(\params) = \mathcal{L}\left(\psi_\params\right),
\end{equation}
where $\psi_\params$ is the wavefunction of the quantum state prepared by the PQC. Unlike on classical hardware, one does not have explicit access to the prepared state. Therefore, the loss $\mathcal{L}(\params)$ needs to be estimated using quantum circuit measurements. PQCs are commonly trained via gradient descent methods, such as finite distance gradients or the \textit{parameter shift rule}~\cite{li2017hybrid,mitarai2018quantum,schuld2019evaluating}, or via gradient-free optimizers such as CMA-ES~\cite{hansen1996adapting}.  \new{For an in-depth introduction to PQCs and VQAs, as well as a broad overview of their potential applications, we refer to Ref.~\cite{Cerezo2021_vqareview}
.}
\subsubsection{Quantum Circuit Born Machines}\label{ssec:methods_QCBM}
\textit{Quantum circuit Born machines} (QCBMs)~\cite{Benedetti2019} are quantum models for solving generative learning tasks, which encode probability distributions over binary data~\footnote{QCBMs can model probability distributions over arbitrary discrete data, but we restrict our attention to the case of binary random variables. Given the possibility of encoding discrete random variables as random bitstrings, this entails no loss of generality.} as measurement probabilities of a wavefunction prepared by a PQC. The probability assigned to a binary string \textbf{x} by a QCBM with circuit parameters $\params$ is given by the Born rule,
\begin{equation}\label{eq:born_probabilities}
    q_\params(\textbf{x}) = |\braket{\textbf{x}}{\psi_\params}|^2,
\end{equation}
where the parametrized wavefunction $\psi_\params$ encodes the distribution $q_{\params}(\textbf{x})$. QCBMs are capable of representing complicated probability distributions~\cite{du2018expressive,Glasser2019,sweke2020learnability, Coyle2019, hinsche2021learnability}, while still permitting a direct means of generating samples from any learned distribution by measuring the associated wavefunction $\psi_\params$. However, much is still unknown about the performance of QCBMs on near- to mid-term quantum devices, especially when modeling complex real-world datasets~\cite{Benedetti2021,Gili2022,Hinsche2022}.

Many methods exist for training a QCBM to minimize a problem-specific loss function, which depends on a dataset $\mathcal{D}$ of size $|\mathcal{D}|$ and circuit parameters $\params$. The loss function we use here is the Kullback-Leibler (KL) divergence between the QCBM distribution $q_\params$ and the evenly weighted empirical distribution $p_\mathcal{D}$ associated to $\mathcal{D}$, given by
\begin{equation}\label{eq:KL_divergence}
\begin{aligned}
    \mathcal{L}\left(\params\right) = \text{KL}\left(p_\mathcal{D} || q_\params \right) & = \mathbb{E}_{\textbf{x}\sim p_\mathcal{D}(\textbf{x})}\left[\log \frac{p_\mathcal{D}(\textbf{x})}{q_\params(\textbf{x})} \right]\\
    & = - \log\left(|\mathcal{D}|\right) - \frac{1}{|\mathcal{D}|} \sum_{\textbf{x} \in \mathcal{D}} \log\left(q_\params(\textbf{x})\right).
\end{aligned}
\end{equation}
For non-uniform weighting, the last line of Eq.~\ref{eq:KL_divergence} must be replaced by the appropriate expectation $\mathbb{E}_{\textbf{x}\sim p_\mathcal{D}(\textbf{x})}$.

\subsubsection{Variational Quantum Eigensolver}\label{ssec:methods_VQE}
The variational quantum eigensolver (VQE)~\cite{peruzzo2014VQE} is a prototypical example of a variational quantum algorithm. The goal in VQE is to find the ground state $\psi_0$ or the ground state energy $E_0$ of a Hamiltonian $H$, which can be found by minimizing the variational energy function
\begin{equation}\label{eq:vqe_energy}
    \mathcal{L}(\params) = E(\params) = \langle \psi_\params | H | \psi_\params \rangle
\end{equation}
of the parametrized trial wavefunction $\psi_\params$ on a quantum computer. \new{This is done by sampling $\psi_\params$ in multiple bases to estimate the expectation values of each operator in the Hamiltonian $H$ with finite precision.} The VQE algorithm can be used to calculate important properties of Hamiltonians in domains of significant practical interest, for example computing the energy of a molecule in the setting of quantum chemistry. In this setting, the qubit Hamiltonian is obtained from the fermionic Hamiltonian of the participating electrons using, for example, the Jordan-Wigner transformation~\cite{somma2002simulating}. Given the practical nature of the problem, and the decades of classical computational techniques towards solving such high-value problems, gave rise to highly specific quantum circuit ansätze and parameter initialization~\cite{romero2019strategies,wecker2015progress}.

\subsection{Tensor Networks}\label{ssec:methods_TN}

\textit{Tensor networks} (TNs) are linear-algebraic models first developed for \new{representing and classically} simulating \new{statistical models} and complex many-body quantum systems~\cite{orus2014practical}, \new{but they} have more recently also been employed as machine learning models~\cite{stoudenmire2016supervised, Novikov_2016, han2018unsupervised, Huggins_2019}.
\new{Tensors are generalizations of vectors and matrices to higher dimensions. The number of axes in a tensor is often called its \textit{order}, where order-1 and order-2 tensors represent vectors and matrices, respectively. A $N$-qubit wave function is arguably most naturally represented by an order-$N$ tensor where every axis has dimension $2$. One approach to reduce the complexity of handling these large tensors with exponentially many entries is to factorize them into a network of lower-order tensors, which, when multiplied together (commonly called \textit{contracted}), recover the original wave function. Depending on the dimension of the axis resulting from the factorization, the tensors can be efficiently stored and used for computation.}

The manner in which \new{the tensors} are contracted together is determined by an undirected graph, with different graphs determining different TN architectures. The nodes of each such graph correspond to \textit{cores} of the TN, while the edges \new{correspond to \textit{indices} or \textit{links} of the TN}, describ\new{ing} tensor contractions to be carried out between pairs of tensor cores \new{along one or more links}. For applications to quantum simulation, the number of nodes $N$ in a TN can, for example, be equal to the number of qubits in the quantum computer, and the topology of the $N$-node graph determines the forms of entanglement which can be faithfully reproduced in the classical TN simulation.

In this work, we utilize \textit{matrix product states} (MPS), computationally tractable TN models whose \new{tensors} are connected along a line graph \new{(Fig. \ref{fig:carton_example}). The tensors are order-3 tensors in the bulk and order-2 tensors at the boundary. Each tensor contains a physical index representing the qubit, and so-called virtual link that connect to neighboring tensor cores. MPS} have a long history in th\new{e} ground state \new{computation} of quantum 1D spin chains~\cite{affleck1987rigorous, fannes1992finitely} via the \textit{density matrix renormalization group} (DMRG) algorithm~\cite{white1992density}, as well as for the efficient simulation of quantum computers with limited entanglement~\cite{vidal2003efficient}. Despite their simplicity, MPS \new{with} sufficient \new{bond dimension} can simulate any $N$-qubit wavefunction, making them a natural first model for many TN applications. The expressivity of an MPS is determined by its \textit{bond dimension} $\chi$ \new{(i.e., the dimension of the shared link between neighboring tensors)}, a quantity associated to the edges of an MPS which sets an upper bound on the amount of entanglement achievable in a simulated quantum state~\cite{garcia_mps}. 
In cases where the entanglement of a quantum state is greater than an MPS is able to exactly reproduce, the \textit{singular value decomposition} (SVD) \new{may} be used to find \new{a low-rank} MPS approximation of the state with near-optimal fidelity.

Although we focus on MPS, our results can be straightforwardly extended to more complicated TN architectures, allowing for different tradeoffs between expressivity and classical computational complexity~\cite{markov2008simulating}.

\subsubsection*{Tensor Network Born Machines}\label{ssec:methods_TNBM}
One application of MPS here is as \textit{tensor network Born machines} (TNBMs)~\cite{han2018unsupervised}, generative models which represent a probability distribution using a simulated quantum wavefunction parametrized by a TN. As such, they form the tensor network analog to QCBMs described in Sec.~\ref{ssec:methods_QCBM}, and we utilize TNBMs to provide the classical solutions to them.

While QCBMs and TNBMs are similar mathematically, with both model families using the Born rule to parametrize classical probability distributions, they nonetheless have distinct complementary benefits in real-world applications. QCBMs are fully quantum models which are able to leverage advances in quantum hardware to better reproduce the correlations present in complex datasets, but are limited by the state of current noisy intermediate-scale quantum (NISQ) devices. By contrast, TNBMs can take full advantage of recent developments in classical computing hardware, notably the development of powerful graphical/tensor processing units (GPUs/TPUs), but are fundamentally limited in their expressivity by the extent of entanglement they are able to simulate efficiently. Additionally, the analytically explicit construction of TNBMs enables exact calculation of probabilities $q_\params(\textbf{x})$ (see Eq.~\eqref{eq:born_probabilities}) and gradients of a loss function $\mathcal{L}$ with respect to the model parameters $\params$. The complementary strengths of both models naturally motivate the development of hybrid quantum-classical Born machine models, but this is complicated in practice by the difficulty of converting between these models. \new{Throughout this work, we specifically consider TNBMs implemented by 1d MPS, as opposed to general TN structures that this model allows}.

\subsection{MPS to PQC mapping}\label{ssec:methods_mapping}
The parameters of PQCs and TNs can in principle be interconverted freely, with the circuit topology of a PQCs itself forming a TN via classical simulation, and with TN canonical forms~\cite{garcia_mps, shi2006classical} facilitating the representation of a TN as a PQC. In practice though, there are several issues that arise with the latter conversion. The quantum ``circuits'' associated with a direct conversion from TNs to PQCs are composed of 
unitary gates acting on multi-level qudits of varying size, whose compilation into gate sets of real-world quantum computers is itself a non-trivial problem (e.g., see \cite{ran2020encoding}).
In the general case of bond dimension of $\chi$, an MPS will be mapped to a quantum circuit containing multi-qubit gates acting on $\left\lceil\log_2(\chi) \right\rceil + 1$ qubits per gate. Much preferred however is a decomposition into two-qubit gates. This is practical for a variety of reasons. For instance, many quantum hardware realizations natively implement two-qubit gates, removing computational overhead in applying multi-qubit operations. Additionally, two-qubit gates can be more sparingly parametrized, in contrast to the exponentially increasing number of parameters needed to fully control multi-qubit gates.
Despite these challenges, we show in the following that the use of an efficient and high-performance conversion method permits MPS of increasing size and complexity to boost the performance of PQCs within several real-world applications.

We use the MPS decomposition protocol developed in Ref.~\cite{rudolph2022MPSdecomposition}, which augments the analytical decomposition method of Ref.~\cite{ran2020encoding} with intertwined constrained classical optimization steps on the circuit unitaries. Using this protocol, transferring the MPS to a PQC results in $k$ layers of \sufour unitaries with a next-neighbor topology, also called \new{\textit{linear} or \textit{staircase}} layers. \new{We note that this decomposition is performed fully on classical hardware}. The choice of an appropriate \new{value for} $k$ is a hyperparameter of the decomposition, and the quality of the decomposition for a fixed $k$ is limited by the entanglement present in the MPS. Fortunately, the decomposition protocol used in this work allows for sequential growing of the circuit up to a desired fidelity. We refer to Appendix~\ref{apx:MPS_decomposition} for a more detailed description of the decomposition protocol used throughout this work. 

One may wonder how this process is efficient on classical hardware. This is the case because the created linear quantum circuit layers tend to result in the MPS having a lower bond dimension than before. Generally speaking, if the MPS was computationally feasible beforehand, it should also be feasible to decompose it via this technique. This is opposed to alternative approaches of brute-force optimization of the linear layers. In such cases, the intermediate states reached during optimization are not guaranteed to represent an MPS with $\chi_{max}$ equal to or less than that of the original MPS.

To have a chance at improving the previously found MPS results, one needs to extend the linear layers with additional gates that would have been infeasible to simulate classically, i.e., the bond dimension $\chi$ of the MPS would need to be increased, which is likely not possible at a point where one is planning to continue optimization on a quantum computer. Extending the quantum circuit can either come in the form of increased circuit depth, more flexible entangling topologies, or both. In our work, from the $k$ linear layers, we extend only the final layer of \sufour gates to an all-to-all topology\new{, that is, a layer containing \sufour gates between all pairs of qubits. The free parameters of those additional gates are drawn from a normal distribution with zero mean and small standard deviation to not significantly alter the mapped quantum state}. Notably, we then train all existing gates in the circuit, and not just the additional gates. We refer to Appendix~\ref{apx:details_SU4} for details on the \sufour gate circuit ansatz \new{as well as the possible decomposition of such gates into Pauli-gates~\eqref{eq:su4_decomposition}}, as well as \new{to} Appendix~\ref{apx:circuit_extension} for a brief study of the effect of adding additional gates to the mapped MPS quantum circuits.

\begin{figure*}
    \centering
    \includegraphics[width=0.99\linewidth]{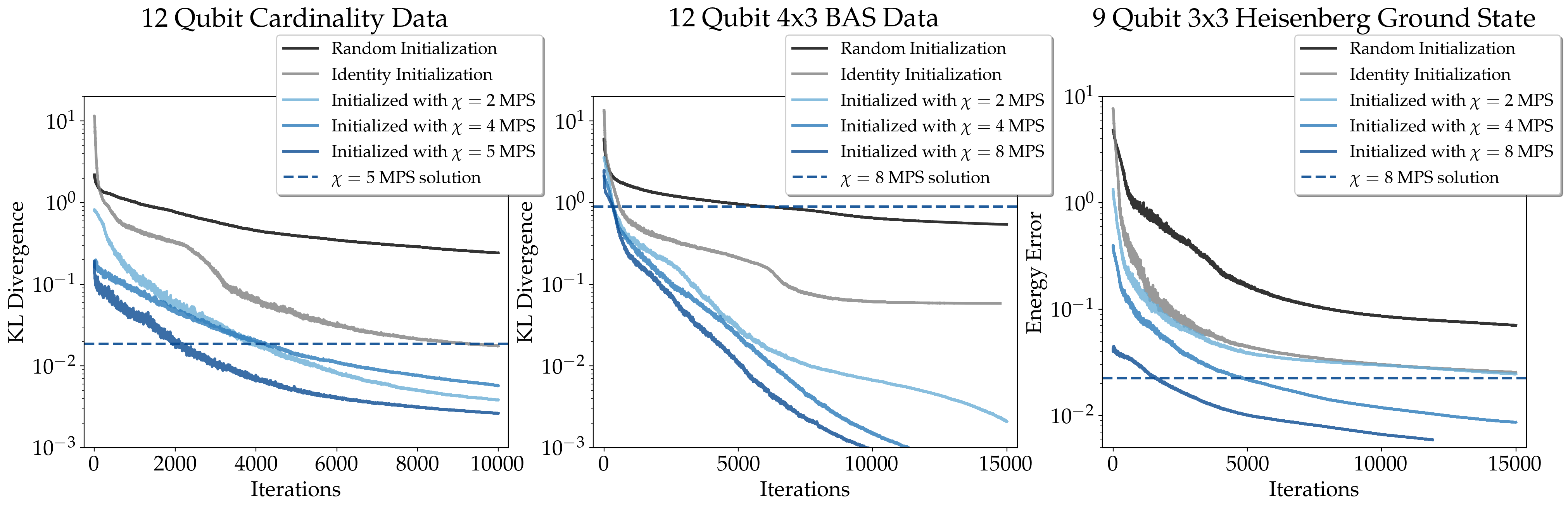}
    \caption{Optimization results for QCBM and VQE training, where each curve represents the best performance among 6 independently initialized runs of the model. The QCBMs are trained to minimize KL divergence relative to datasets of length $N = 12$ strings of fixed cardinality $N / 2$ or $4 \times 3$ bars and stripes images, whereas the VQE optimization aims to minimize the energy of a 2D Heisenberg Hamiltonian with 9 qubits, and with size $3 \times 3$. The quantum circuits for each case consist of three, four, and four layers, respectively, with the \sufour gates of each layer arranged in a linear topology, except for the final layer whose gates are connected in an all-to-all manner. In each case, quantum circuits whose parameters are initialized randomly or close to the identity exhibit worse final losses than those initialized with a classically trained MPS model. Additionally, the use of increased classical resources (as quantified by the bond dimension $\chi$) leads to improved performance of the trained quantum circuits. All optimization runs compared inside individual plots share exactly the same circuit layout and number of trainable parameters. Differences in training performance are only due to different initial parameters.}
    \label{fig:training_results}
\end{figure*}

\section{Results and Discussion}~\label{sec:results}


To assess the benefits of the synergistic training framework for real-world applications, we explore the performance of this method in a variety of \new{generative} modeling tasks\new{, where the goal is to learn to reproduce the discrete measurement outcome distribution that is given by the training data,} and \new{a} Hamiltonian ground state search problem. \new{The corresponding MPS minimize the same loss function as the following PQCs, i.e., the KL divergence~\eqref{eq:KL_divergence} for the generative modeling tasks, and the energy of the Hamiltonian for the ground state search.} In each case, we compare quantum circuits trained using our MPS initialization approach to those which are initialized randomly or with all gates close to the identity. The latter has been shown empirically to reduce the effects of barren plateaus and improve convergence behavior at the start of PQC optimization~\cite{grant2019initialization}. 

We study the impact of different circuit architectures on our results by designing the quantum circuit layers with either linear or all-to-all topologies of fully parametrized \sufour gates. While all-to-all connectivity is likely not practical in a scalable manner on near- to mid-term quantum hardware, it provides a challenging use case for an initialization method leveraging a TN model with linear connectivity, while also illustrating an important advantage that quantum hardware has over classical TN simulation techniques: The flexible choice of circuit depth and entangling topology. Implementation details can be found in Appendix~\ref{apx:details}.

Our results show that the use of TNs as \new{a} strategy \new{to initialize the parameters of quantum circuits} succeeds in boosting the performance of \new{PQCs} in all of these tasks, with an increase in classical computing resources (as quantified by the bond dimension $\chi$ of the MPS) in nearly every case leading to a corresponding increase in the final performance of the trained quantum circuit. This is reflected not only in the final losses in different tasks, but also through an analysis of the parameter gradients seen by the circuit at initialization. We find that although randomly initialized quantum circuits exhibit gradients of exponentially vanishing magnitude in system size, a manifestation of barren plateaus within generative modeling, the use of classically trained MPS to \new{provide learned} initialization avoids this phenomenon entirely.

In our first experiments, we explore the optimization performance of QCBM and VQE, i.e., the progression of the loss function values (Eqs.~\eqref{eq:KL_divergence} \& \eqref{eq:vqe_energy}, respectively).
For the QCBM \new{and its TN equivalent, the TNBM}, we study two distinct datasets of bitstrings of length $N = 12$. The first is the \textit{cardinality} dataset that is the set of all strings having a cardinality (i.e., Hamming weight, or number of 1s) of $N / 2$. The second QCBM dataset is the dataset of bars and stripes (BAS) images~\new{\cite{MacKay-book-2002,benedetti2019parameterized}} containing horizontal or vertical lines \new{on a 2D pixel layout. The Cardinality dataset is a dataset with moderately low correlations between neighboring bits, whereas the BAS dataset is a dataset which exhibits strong correlations between bits within the same row or column, and thus makes it a 2d-correlated dataset.} The VQE optimization problem uses $N = 9$ qubits and minimizes the energy of the 2D Heisenberg model Hamiltonian on a $3 \times 3$ rectangular lattice:
\begin{equation}
\label{eq:heisenberg_hamiltonian}
  H = \frac{1}{4} \sum_{\langle i, j \rangle} \sigma_{X}^{(i)} \sigma_{X}^{(j)} + \sigma_{Y}^{(i)} \sigma_{Y}^{(j)} + \sigma_{Z}^{(i)} \sigma_{Z}^{(j)}.
\end{equation}
$\langle i, j \rangle$ indexes all nearest-neighbor spins in a 2D rectangular grid with open boundary conditions, and $\sigma_{\mu}^{(i)}$, $\mu = X, Y, Z$ denote the Pauli operators acting on the $i$'th spin. We measure the energy error $\Delta E(\params) = E(\params) - E_0$ relative to the exact ground state energy $E_0$

In all cases, we compare the performance of PQCs initialized with random \sufour or near-identity unitaries to those initialized with previously found MPS solutions. Transferring the MPS state is done via the decomposition protocol described in Ref.~\cite{rudolph2022MPSdecomposition}. As described in Sec.~\ref{ssec:methods_mapping}, the topologies utilized here contain $k$ layers of gates, which are arranged linearly in the first $k-1$ layers and in an all-to-all topology in the last layer. For the cardinality dataset we utilize $k=3$ layers, and for the BAS dataset, as well as for the 2D Heisenberg Hamiltonian, $k=4$ layers. The parameters of the quantum circuits are optimized using the \textit{CMA-ES} algorithm~\cite{hansen1996adapting, hansen2019pycma}, a gradient-free optimizer that is based on an adaptive evolutionary strategy.

Our optimization results in Fig.~\ref{fig:training_results} depict the best optimization runs out of $6$ repetitions\new{, i.e. the runs that reach the lowest loss after the prescribed training iterations}. It becomes evident that the models without the MPS initialization do not converge to high-quality solutions. In fact, we have observed that, while all-to-all layers clearly increase the expressive capabilities of a PQC as compared to linear layers, the presence of a single all-to-all entangling layer has detrimental effect on their trainability (see also Appendix~\ref{apx:circuit_extension}). By choosing an initialization which makes use of the parameters of a classically trained MPS model however, all models exhibit a drastic increase in performance on all the tasks we considered. This behavior is enhanced even more by the use of MPS with larger bond dimension $\chi$. We emphasize that all PQC models compared inside one plot have precisely the same circuit layout and number of parameters. Differences in performance are only due to different choices of parameter initialization.

\begin{figure*}[t]
    \centering
    \hspace{1cm}\includegraphics[width=0.90\linewidth]{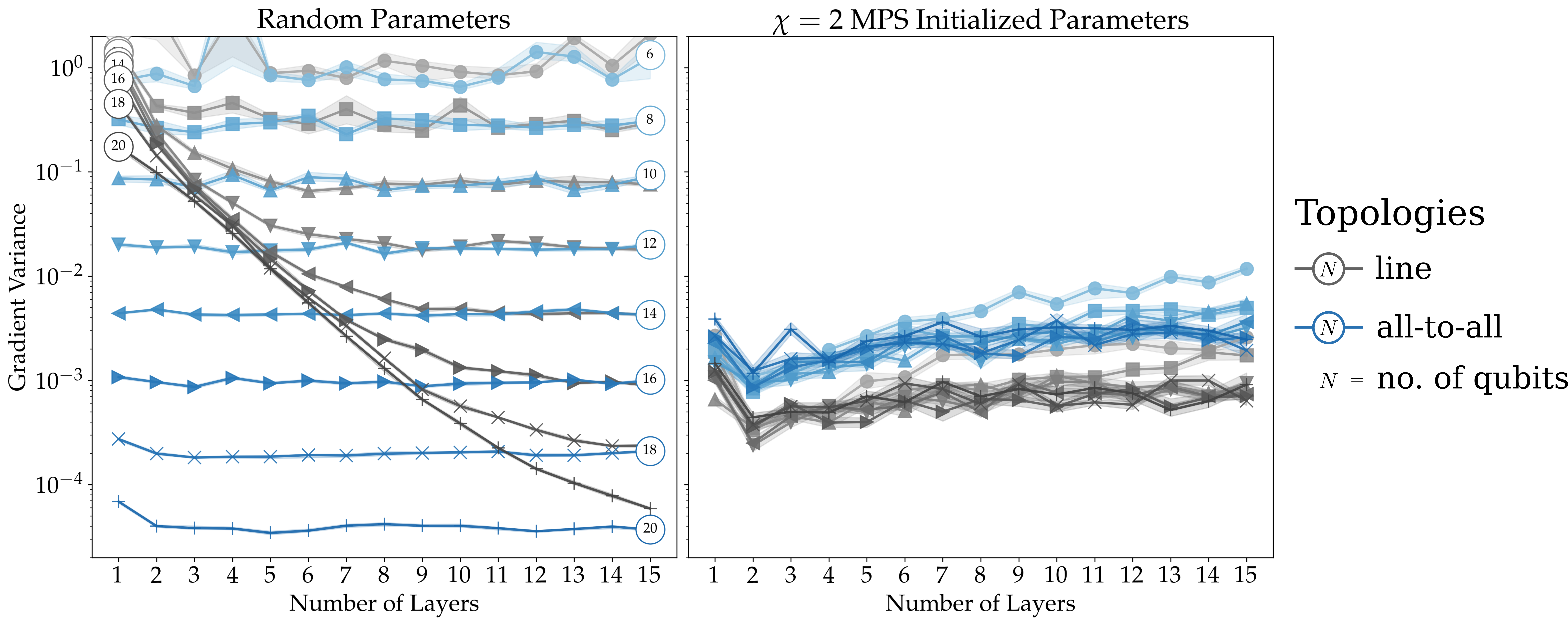}
    \caption{\new{Prevention} of barren plateaus in QCBMs inside our synergistic optimization framework as demonstrated by the gradient variances with respect to the KL divergence loss in Eq.~\eqref{eq:KL_divergence}. The grey lines indicate the gradient variances of linear topology circuits, whereas the blue lines indicate the gradient variances of all-to-all topology circuits. The numbers at the beginning or the end of the lines denote the number of qubits. We record the median gradient variances  over $1000$ repetitions for the randomly initialized circuits~(left), and $100$ repetitions for the $\chi=2$ MPS initialized circuits~(right), as well as bootstrapped 25-75 percentile confidence intervals of the median inside the shaded areas. We study the Cardinality $N/2$ dataset for the respective number of qubits $N$. The gradients are measured with respect to the YY-entangling gate contribution of the first \sufour gate in the circuit between qubit $1$ and $2$. For random parameter initializations, the gradient variance decays exponentially in the number of qubits, and also the circuit depth up until a certain limit. This is clear indication for the existence of barren plateaus. One all-to-all layer of \sufour gates appears to fully maximize the degree of barrenness. In contrast, the gradient variances of MPS initialized circuits neither decay significantly in the number of qubits nor with increasing circuit depth.}
    \label{fig:gradient_results}
\end{figure*}

One may expect that the informed initialization merely affects the number of quantum circuit evaluations needed to achieve a target training loss, however, we show that it qualitatively changes convergence behavior. For example, in the case of the cardinality dataset, QCBMs initialized with $\chi=4$ or $\chi=5$ \new{TNBMs} begin training at approximately the same KL values, but the PQC initialized with the larger $\chi$ \new{TNBM} rapidly achieves better loss values shortly after. 
\new{S}imilar behavior can be seen in the case of the more challenging BAS dataset. In this case, the MPS solutions achieve relatively high KL divergence values, which consequently leads to high initial KL values for the MPS\new{-}initialized QCBMs. While the randomly initialized circuits generally reach the KLs at which the MPS initialized models start out, the latter are able to converge fully, while the former appear to plateau at much higher KL values.

Crucially, we observe in the VQE optimization example that initializing with $\chi=2$ MPS does not suffice to reliably improve over a naive near-identity initialization of the circuit unitaries. Only MPS with larger bond dimension $\chi$, facilitated by the layer-efficient decomposition in Ref.~\cite{rudolph2022MPSdecomposition}, enable significant enhancements. This is also highlighted by the depicted loss values achieved by the MPS solutions with highest $\chi$ that were used to initialize the respective PQC models. In the VQE simulation, initializing the PQC with $\chi=2$ is not sufficient for the PQC to outperform what the MPS on classical hardware may have been capable of. The particular case of $\chi=2$, where the MPS readily maps to two-qubit gates, was studied in Ref.~\cite{dborin2022pretraining}, and does not require the layer-efficient decomposition scheme used in this work which enables arbitrary $\chi$. 

However, the final MPS losses (indicated by the dashed lines) also showcase how the PQC solutions can improve on solutions attained on classical hardware by leveraging the more flexible capabilities of quantum hardware and initializing with strong classical models. The gaps between the final MPS losses and the respective PQC initializations stem from imperfect decomposition of the MPS into a low number of two-qubit gate layers, as well as the close-to-identity extension of the quantum circuits into the all-to-all topologies, and the initial exploration step size of the CMA-ES optimizer. 
While initializing of the QCBM with \new{a} $\chi=8$ \new{TNBM} on the BAS dataset here achieves the best result, we note that it does not clearly outperform $\chi=4$ on average (see Fig.~\ref{fig:median_training_results} in Appendix~\ref{apx:median_results})
The likely reason is that the BAS dataset is 2\new{D}-correlated and thus the MPS with growing bond dimension $\chi$ increasingly biases the quantum circuit to a 1d-correlated solution. In other words, there is a bias mismatch between the TN architecture used and the task at hand. Depending on the number of additional free parameters that the PQC is given access to, this can lead to saddle points and local minima, because the PQC needs to correct the unsuitable bias. In such cases, one may try to train another TN model which is adapted for more general correlation structures~\cite{Lei_VQE_PEPS}, and then, if needed, map this TN to a quantum circuit \cite{Ian_PEPS_to_QC,Lei_VQE_PEPS}. \new{Future work will need to study how to best extend pretrained quantum circuits with additional gates, i.e., where to most efficiently place additional gates such that the PQC can improve on the TN solution and potentially escape its bias.}

To assess whether the synergistic framework is expected to be effective at improving the trainability of PQCs as the number of qubits increases, we now assess the variance of parameter gradients, i.e. the \textit{barrenness}, of QCBMs training on the cardinality dataset. The results are shown in Fig.~\ref{fig:gradient_results}. We probe the gradient of the KL divergence loss with respect to the parameter controlling the YY-entangling component (according to the KAK-decomposition~\cite{tucci2005kak}) of the first \sufour gate between qubits $1$ and $2$ (see Appendix~\ref{apx:details_SU4} for details). Gradient magnitudes for that parameter are recorded $1000$ times per data point in the case of random parameters, and $100$ times per data point in the case of \new{the TNBM initialization with} $\chi=2$. The latter case contains the training of the MPS, as well as the mapping to a quantum circuit, and the (potential) extension of the linear layers to all-to-all topologies. We note that our results are robust to different choices of the parameter for which the gradients are estimated.

In the case of randomly initialized parameters ($\params \in [0, 2\pi]$), we observe a clear exponential decay in gradient variances with increasing circuit depth and number of qubits. To the best of our knowledge, we provide the first experimental demonstration of barren plateaus for QCBMs trained with the KL divergence loss function. The nature of this decay depends on the quantum circuit topology used, with a single all-to-all layer being sufficient to saturate the barreness for a specific number of qubits up until $N=18-20$. In contrast, we observe that QCBMs initialized with a classically trained MPS avoid this exponential decay\new{--} something which likely contributes to the increased trainability observed in Fig.~\ref{fig:training_results}. Fascinatingly, the gradients in this initialization can actually exhibit an \textit{increase} in variance with circuit depth, a trend which is more visible in the all-to-all extended circuit. Overall, the gradients of the all-to-all topologies have a larger gradient variance, indicating that the circuit extension after transferring the MPS solution is crucial to the success of the PQC. These findings show that the use of trained MPS places the quantum circuit model in a region of the parameter space without evident barren plateaus, but where the additional flexibility provided by increased connectivity in the quantum circuit enables it to effectively improve on the classical MPS solution. With a more sparse set of measurements, we can confirm a very similar trend when utilizing $\chi=4$ MPS solutions. 

\begin{figure}[t]
    \centering
    \includegraphics[width=0.98\linewidth]{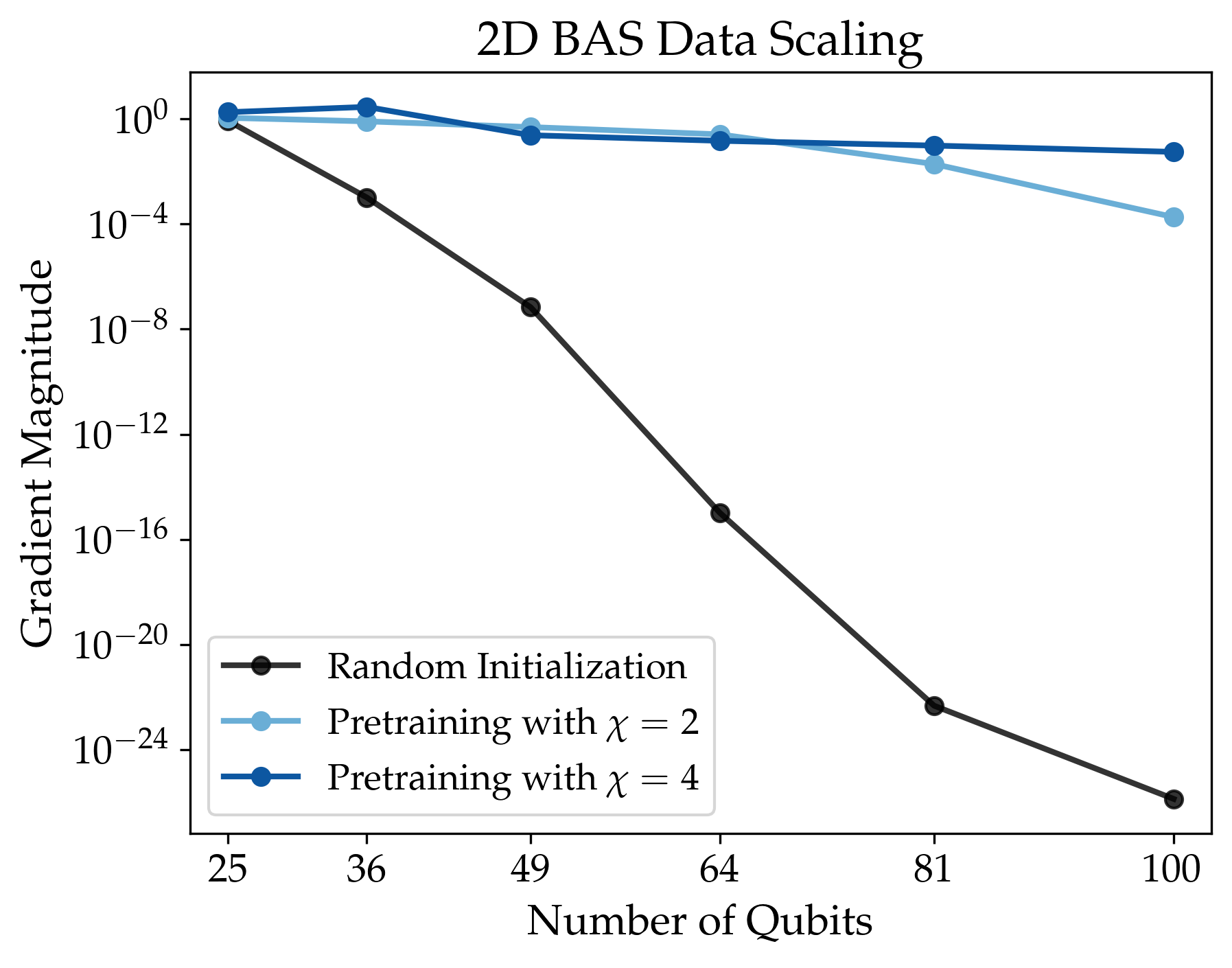}
    \caption{\new{Gradient magnitude scaling for a QCBM with the KL divergence loss function and the BAS dataset. For the pretrained cases, we train MPS with bond dimensions $\chi=2$ or $\chi=4$, decompose them into one layer of \sufour gates while optimizing the fidelity, and extend that layer to a 2D topology using identity gates. The gradient magnitude, i.e., the 2-norm of the gradient vector, is then evaluated on an MPS-based quantum circuit simulator for practical feasibility. While the gradient magnitude of the randomly initialized circuits decay exponentially with the number of qubits, the pretrained cases exhibit significantly more stable behavior. After $9\times 9=81$ qubits, the gradients for the $\chi=2$ pretraining start to decay and are surpassed by the $\chi=4$ case.}}
    \label{fig:bas_gradient_scaling}
\end{figure}

\new{Several potential criticisms may be raised about the scenario studied above and presented in Fig.~\ref{fig:gradient_results}. First, while statevector simulation allows us to generate valuable statistics for deep circuits, it only permits us to consider system sizes and datasets up to 20 qubits. This limitation is particularly restrictive when attempting to highlight the scalability of our method since trainability issues induced by barren plateaus are expected to manifest themselves more prominently as the qubit count increases. Consequently, we had to extend the decomposed circuits into an all-to-all topology to showcase the utility of our method more discernibly at such a limited qubit count. This is a second potential criticism because the study of all-to-all topologies is unlikely to be highly relevant in practice given the sheer number of noisy gates and possibly restricted hardware connectivity. Finally, the correlation structure in the Cardinality dataset is such that an MPS with bond dimension $\chi$ linear in the number of qubits can exactly represent the target distribution. Consequently, one might expect pretraining using an MPS to be abnormally successful. This fear is only partially supported by our findings in Fig.~\ref{fig:training_results} because, while initial losses after pretraining on the BAS dataset are high, the resulting QCBM optimization is most dramatically improved.}

\new{We aim to address all these potential concerns with a complementary gradient scaling result using MPS-based quantum circuit simulation and a generative modeling task on the BAS dataset in a square arrangement. The 2D correlations in the BAS dataset suggest that a favorable circuit ansatz for a QCBM is one comprised of \sufour gates in a 2D next-neighbor topology. Notably, this resembles a practical circuit topology for which quantum devices could exhibit an advantage, given the hardness of many 2D problems and the hardware connectivities in various modern quantum devices.}

\new{For the benchmarks, we train $N$-qubit TNBMs with $\chi=2$ and $\chi=4$ on all $\mathcal{O}(2^{\sqrt{n}})$ samples from the $\sqrt{n}\times \sqrt{n}$ BAS dataset. We then decompose the corresponding MPS into one linear layer of \sufour gates and extend that layer into a 2D topology using identity-initialized \sufour gates.
For the random quantum circuit reference case, the linear part of the topology is randomly initialized, but the extension to the 2D topology is again done using identity operations. The gradients are computed via automatic differentiation of the MPS-based quantum circuit simulator. The identity initialization of the additional gates helps us simultaneously keep the gradient computations both feasible and exact by avoiding the need for the truncation of the simulator MPS.} 

\new{Fig.~\ref{fig:bas_gradient_scaling} depicts the scaling of the gradient magnitude of the KL divergence loss function with respect to the circuit parameters, i.e., the 2-norm of the gradient vector, up to $10\times 10 = 100$ qubits. Even in this new numerical setup, we observe results that are exactly consistent with the results in Fig.~\ref{fig:gradient_results} for the $\chi=2$ case, but we are now able to see that pretraining using a $\chi=4$ eventually outperforms and keeps up the favorable scaling. This confirms the intuition that increasing classical resources are required as the problem size increases, and that high-performance schemes to convert tensor network states into quantum circuits will be needed in the future. However, it also suggests that moderate classical resources are sufficient in order to continue to provide value for the following quantum circuit optimization. One may have expected that drastic increases in classical compute would be required to escape barren plateaus, but we show that this is in fact not exponentially demanding using a synergistic framework jointly leveraging TNs and PQCs.}

The \new{avoidance} of barren plateaus, as indicated in Fig\new{s}.~\ref{fig:gradient_results} \new{and~\ref{fig:bas_gradient_scaling}}, is vital to ensuring the trainability of PQCs and their viability on quantum hardware. Vanishing gradient variances imply that gradient magnitudes also vanish~\cite{Mcclear2018Barren}, which leads the estimation of parameter gradients on quantum hardware to require a number of measurements which grows exponentially in the number of qubits. Additionally, barren plateaus have been linked to the phenomena of \textit{cost concentration} and \textit{narrow gorges}~\cite{arrasmith2021gorges}, which hinder the ability of gradient-based and gradient-free optimizers to find high-quality solutions, as well as the existence of large numbers of low-quality local minima~\cite{anschuetz2021critical}, which present further difficulties in learning.  
Aside from improving the training performance in practice (as seen in Fig.~\ref{fig:training_results}), stable gradient variances (such as those in Fig\new{s}.~\ref{fig:gradient_results} \new{and~\ref{fig:bas_gradient_scaling}}) hint that a finite (or at worst, non-exponential) number of quantum circuit evaluations may be sufficient to estimate parameter gradients and perform PQC optimization on quantum hardware in a scalable manner.


\section{Conclusions}\label{sec:conclusions}

This work introduces a synergistic training framework for quantum algorithms, which employs classical tensor network simulations towards boosting the performance of PQCs. Our framework allows a problem of interest to be attacked first with the aid of abundant classical resources (e.g. GPUs and TPUs), before being transitioned onto quantum hardware to find a solution with further improved performance. By moving the work of quantum computers to improv\new{e} on promising classical solutions, rather than finding such solutions \textit{de novo}, we ensure that scarce quantum resources are allocated where they are most effective, setting up parametrized quantum algorithms for success.

Assessing the performance of our methods on generative modeling and Hamiltonian minimization problems, we found that PQCs initialized with this synergistic training framework not only obtained better training losses using identical quantum resources, but also exhibited qualitatively improved optimization behavior, with deep quantum circuits transformed from being practically untrainable to reliably converging on high-quality solutions. A study of gradient variances \new{and magnitudes} shows the promise of this method for avoiding barren plateaus and related worst-case guarantees, which for randomly initialized PQCs lead to gradients which decay exponentially in both the number of qubits and the depth of the circuit. For PQCs initialized using classically obtained TN solutions however, we observed gradient variances \new{and magnitudes} which remain essentially constant with respect to size, even showing a slight increase in deeper circuits. \new{At system sizes up to 100 qubits, we witnessed a change of trends that favored pretraining with larger bond dimensions in order to keep up the favorable scaling. Nevertheless, we demonstrate that classical computing resources do not need to drastically increase in order to keep mitigating barren plateaus in PQCs to a very strong degree.} These results point towards the promise of this framework for enabling PQCs to scale to large number of qubits, thereby unlocking the latent capabilities of quantum computers for optimization and learning problems which remain out of reach for purely classical methods.

These findings naturally open up several related questions. We have employed MPS as our classical TN ansatz, whose bond dimension $\chi$ determines the classical resources allocated for PQC initialization, but have yet to characterize the performance of our method on problem sizes requiring very large values of $\chi$. While our findings show that larger bond dimensions in our synergistic framework lead to increased PQC performance, we also anticipate an eventual need to employ more sophisticated TN models whose topology is better adapted to the connectivity of the circuit architecture at hand. To this end, initializing PQCs using tree tensor networks is a natural next area of study, as the simple canonical forms available to such models permit a straightforward extension of the decomposition procedure used here~\cite{rudolph2022MPSdecomposition}. We anticipate the use of more flexible TN models to lead to further improvements in the performance of quantum algorithms, complementary to those identified for the use of larger values of $\chi$.

As a final remark, we note that our synergistic framework highlights the benefits of moving beyond the adversarial mindset of ``classical vs. quantum'' which is typical of discussions surrounding quantum supremacy. By embracing the rich connections between classical TN algorithms and PQCs, we show that good use can be made of the complementary strengths of both. Moving forward, we believe that the existence of powerful classical simulation methods should not be seen as an obstacle on the path to demonstrating practical quantum advantage, but rather as a guide to help quantum methods find the\new{ir} way~\cite{PerdomoOrtiz2017}.



\begin{acknowledgments} 

The authors would like to acknowledge Vladimir Vargas-Calderón, Brian Dellabetta, Dmitri Iouchtchenko, Peter Johnson, Yudong Cao, Kaitlin Gilli, Dax Enshan Koh, and Mohamed Hibat-Allah for their helpful feedback on early versions of this manuscript. The authors would like to acknowledge the \orquestra platform by Zapata Computing Inc. that was used for collecting the data presented in this work.

\end{acknowledgments}


\section*{DATA AVAILABILITY}
All data needed to evaluate the conclusions of this work are available from the corresponding author upon reasonable request.

\section*{AUTHOR CONTRIBUTIONS}
M.S.R. and A.P.O. conceived the initial proposal for the synergistic framework. J.M. contributed to the interpretation of the synergistic framework in its final form. M.S.R. wrote the code used in this work and performed most numerical simulations. \new{D.M. performed the numerical simulations for Fig.~\ref{fig:bas_gradient_scaling}}. J.C. provided optimized MPS models for the VQE simulations. J.M., J.C. and  A.A. provided relevant expertise in tensor network methods. A.P-O helped supervise and coordinate the efforts in this work. All authors regularly analyzed the numerical results and contributed to the final version of the manuscript.

%

\appendix

\section{Distribution of gradient magnitudes}
\begin{figure*}
    \centering
    \includegraphics[width=0.95\linewidth]{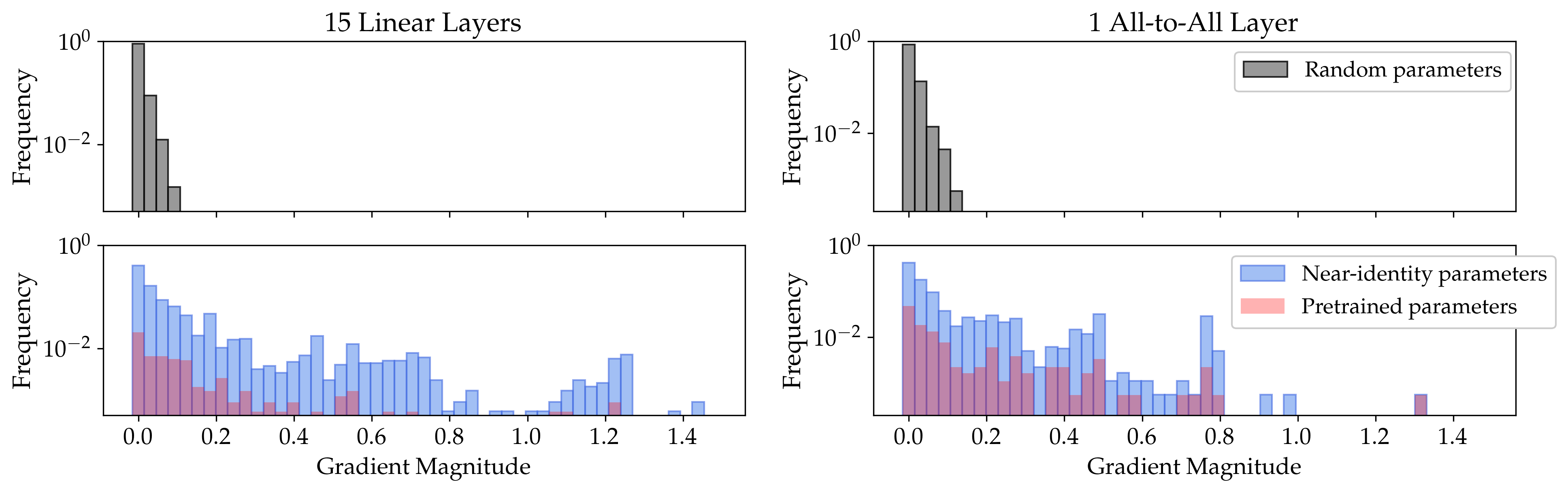}
    \caption{\new{Distribution of gradient magnitudes in the gradient vectors of individual 16-qubit QCBMs with random parameters or pretrained parameters using a $\chi=2$ TNBM. The circuits consist of 15 linear layers (left) or one all-to-all layer (right) of \sufour gates, where only the linear component of the first layer was pretrained.}}
    \label{fig:gradient_magnitudes}
\end{figure*}

\new{To supplement the expected gradient variance statistics depicted in Fig.~\ref{fig:gradient_results}, we now present a more fine-grained study of the gradients within individual models. For this, we pick the 16-qubit instance of the Cardinality dataset and calculate the entire gradient vector of four QCBMs with respect to the KL divergence loss function~\eqref{eq:KL_divergence}. The circuits are either 15 linear layers, or one all-to-all layer of \sufour gates. We then record the distribution of gradient vector entries at random parameters or pretrained parameters from a $\chi=2$ bond dimension TNBM. The results are shown in Fig.~\ref{fig:gradient_magnitudes}.}

\new{In the case of random initial parameters, it can be seen that the distribution of gradient magnitudes decays exponentially, making large gradients exponentially unlikely within one model. This is a pattern that is expected for exponential concentration phenomena in general, and is a form of \textit{probabilistic} exponential concentration~\cite{arrasmith2021effect}.
In contrast, the pretrained QCBMs completely break the trend observed in the random circuits and exhibit generous gradient magnitudes for a significant proportion of their parameters. By highlighting in red the pretrained parameters in Fig.~\ref{fig:gradient_magnitudes}, we determine that there is no clear trend which type of parameters (i.e., pretrained or near-identity initialized) tend to have larger gradients. This speaks to the positive influence of both the pretrained parameters and the additional gates. One may have expected that the pretrained parameters are in a local minimum, or more precisely, a saddle point. This would lead to systematically lower gradient for those parameters. At the same time, we observe that the newly added parameters experience large gradients, highlighting that these parameters will be able to be trained in order to improve the pretrained circuit. We attribute much of this positive behavior to the near-identity initialized with parameters that are drawn from a Gaussian distribution with zero mean and a standard deviation of $0.01$.}

\new{Finally, we note that the variation in gradient distributions is quite large between individual pretrained models, which is consistent with the sizeable expected gradient variance displayed in Fig.~\ref{fig:gradient_results}. The results shown here should therefore merely be understood as representative examples. The exponential decay in gradient magnitudes does however appear to be very robust for the randomly initialized QCBM at its respective number of qubits.}

\section{Performance enhancement when extending linear layers}\label{apx:circuit_extension}

\begin{figure*}
    \centering
    \includegraphics[width=0.85\linewidth]{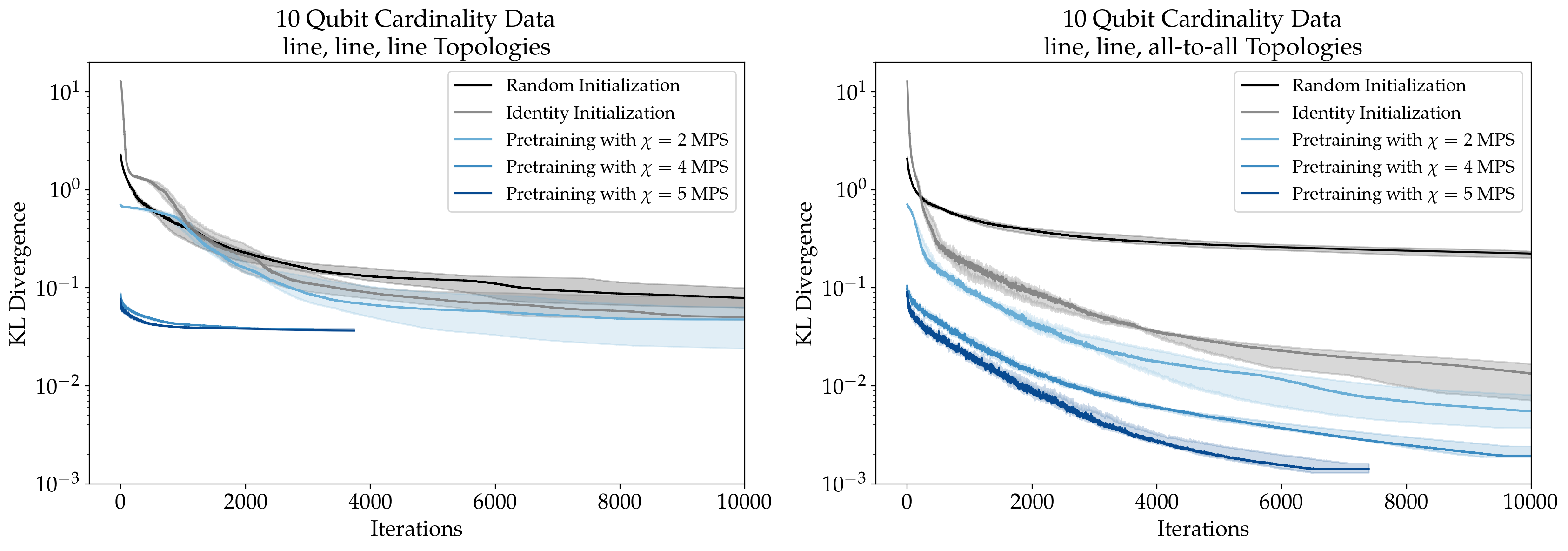}
    \caption{Optimization results for QCBMs training on the Cardinality dataset with $N=10$ qubits and three linear layers. We showcase the enhancement of the mapped MPS solutions when the final linear layer is extended to an all-to-all topology. While trainability suffers for random initial parameters, the classically initialized PQCs are able to converge to significantly better solutions.}
    \label{fig:circuit_extension}
\end{figure*}

When initializing PQCs for model optimization, the approach throughout this work has been to decompose the classically trained MPS into $k$ linear layers of two-qubit gates, and extend only the final layer into an all-to-all topology. \new{This is to say that we add additional two-qubit gates with near-zero parameters to the final linear layer until the every qubit is directly connected to every other qubit. The two-qubit gates used are described in Eq.~\eqref{eq:su4_decomposition}. The precise placement of additional gates here is meant to be generic, and the optimal and most resource-efficient method is likely highly problem dependent.} Fig.~\ref{fig:training_results} demonstrates how \new{PQCs with added gates} can reliably improve on the final loss values that the MPS could achieve. In contrast, Fig.~\ref{fig:circuit_extension} depicts an example case of an $N=10$ qubit QCBM training on the Cardinality dataset where we do not \new{add these additional gates to} the linear layers after transferring the MPS solution to the quantum circuit. Interestingly, with random initial parameters, the linearly arranged circuit reaches better KL divergence values (black, left panel) than when the final layer is extended to an all-to-all topology (black, right panel). This highlights the deterioration of trainability with increasing circuit depth that can also seen in Fig.~\ref{fig:gradient_results}. The near-identity initialization is notably less hampered by this. When classically initializing the QCBMs, the models can only aim to recover the MPS performance when the circuit is not extended with additional gates and free parameters. In contrast, the all-to-all topology allows the QCBM to achieve great performance and, as also seen in Fig.~\ref{fig:training_results}, QCBM initialized with larger $\chi$ MPS perform better even when starting at similar loss values.

\section{Median Optimization Performance from Sec.~\ref{sec:results}}\label{apx:median_results}
\begin{figure*}
    \centering
    \includegraphics[width=0.95\linewidth]{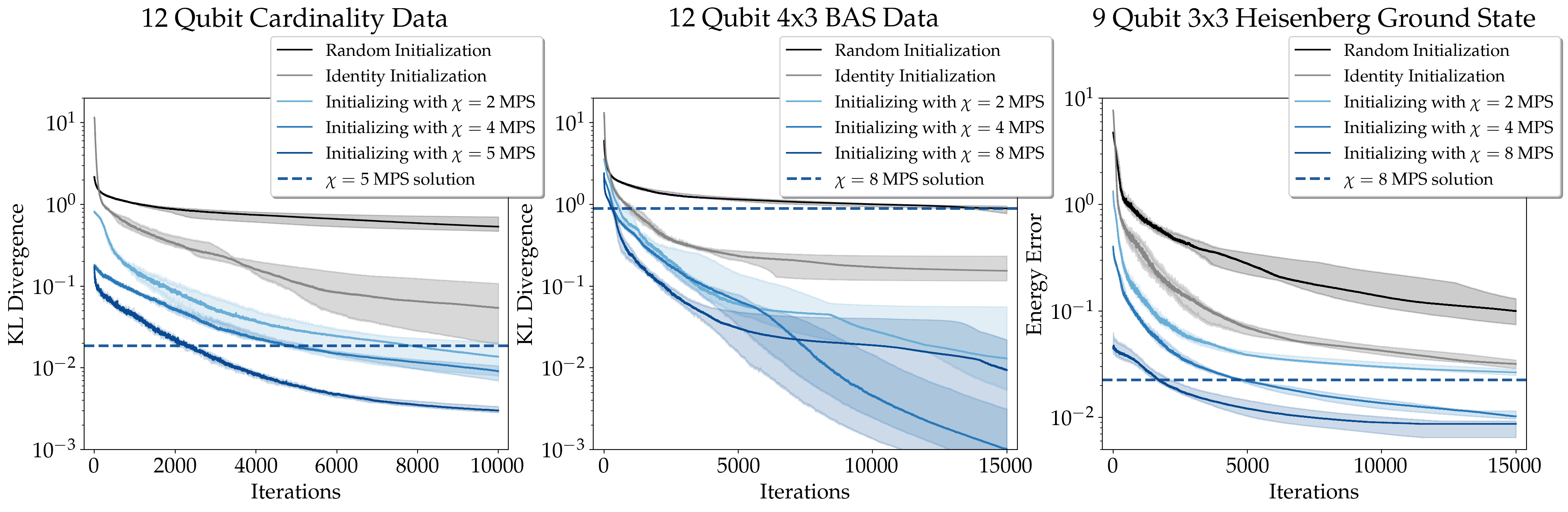}
    \caption{Optimization results for QCBM and VQE training, where each curve represents the bootstrapped median and 25-75 percentiles of 6 independently initialized runs of the model. The data is complementary to the best of $6$ runs shown in Fig.~\ref{fig:training_results}. Initializing the PQCs with classically found MPS solutions appears to be generally beneficial as compared to randomly or near-identity initialized parameters. While continuous improvements can be attained with larger bond dimension $\chi$ in the Cardinality dataset, as well as the Heisenberg ground state minimization, the BAS dataset - which is a 2D correlated dataset - exhibits an unfavorable bias miss-match with the MPS that the PQC needs to un-learn. Even so, the synergistic framework with small $\chi$ greatly improves on the uninformed initializations.}
    \label{fig:median_training_results}
\end{figure*}

In addition to the best of $6$ optimization runs shown in Fig.~\ref{fig:training_results}, we here report \new{the} median and 25-75 percentile of the losses \new{throughout the optimization. The statistics were gathered by bootstrapping, which is a common data re-sampling strategy to enhance the robustness of uncertainty estimates.} Fig.~\ref{fig:median_training_results} shows that initializing the PQC circuits with classically found MPS solutions is beneficial in all cases and significantly enhances the trainability of the models. In the cases of the Cardinality dataset in the generative modeling task, and the Heisenberg ground state minimization, the synergistic optimization framework shows continuous improvements with increasing MPS bond dimension $\chi$. The case of the BAS dataset is different. As discussed in Sec.~\ref{sec:results}, the BAS dataset is a \new{dataset which exhibits strong correlations between bits within the same row or column, which means that it is a} 2d-correlated dataset. \new{On the other hand, the} MPS solutions\new{, due to the line-graph of the TN,} are \new{particularly well-suited to represent} 1d-correlated \new{states}. This introduces an unbeneficial bias miss-match which the PQC after circuit extension needs to un-learn. In such cases, initializing the PQC with large $\chi$ MPS may not deliver the desired robust improvements. Still, the synergistic approach with any $\chi$ greatly outperforms the uninformed random initialization and also the near-identity initialization, which was empirically shown to deal with barren plateaus~\cite{grant2019initialization}.

\section{Implementation Details of the MPS and PQCs}\label{apx:details}

\subsection{MPS optimization}\label{apx:details_MPS}
The MPS for the generative modelling task, i.e., the TNBMs, were implemented and trained according to the method outlined in Ref.~\cite{han2018unsupervised}. The TNBMs are trained via gradient descent using gradients of the KL divergence loss function (see Eq.~\eqref{eq:KL_divergence}) that can be calculated efficiently using a \textit{density matrix renormalization group} (DMRG)~\new{\cite{white1992density}} method. This approach allows for adaptive bond dimensions $\chi$ for each MPS bond up to a maximum of $\chi_{max}$, or a singular value threshold of $5\cdot10^{-5}$ in the SVD truncation. The learning rate $\eta$ for the gradient descent updates was chosen to be $\eta=0.01$, with $50$ optimization sweeps for the results in Fig.~\ref{fig:training_results} for complete convergence, and $30$ optimization sweeps for the gradient results in Fig.~\ref{fig:gradient_results}.

The training of the MPS for Hamiltonian ground state minimization uses an analogous DMRG-based gradient calculation approach for the energy loss function described Eq.~\eqref{eq:vqe_energy}. The MPS calculations utilize the ITensor library~\cite{ITensor_julia}, were $\chi=16$ is used to find the exact ground state without any approximation. The exact MPS was then truncated via SVD to $\chi=2, 4, 8$ for the results in Fig.~\ref{fig:training_results}. This is viable because fidelity with the ground state (which is what SVD empirically optimizes) and the energy in Eq.~\eqref{eq:vqe_energy} are proportional, whereas fidelity is not proportional to the KL divergence in Eq.~\eqref{eq:KL_divergence}.

\subsection{SU(4) gates}\label{apx:details_SU4}
The PQCs throughout this work used \sufour gates between qubit pairs with $15$ parameters per gate. Up to a global phase, \sufour gates represent fully parametrized two-qubit interactions. By the KAK-decomposition~\cite{tucci2005kak}, an \sufour rotation can be decomposed into four single-qubit U$(\new{2})$ rotations \new{(which are fully general single-qubit rotations)}, and the entangling gates, i.e., XX, YY, and ZZ. Concretely,

\begin{align}\label{eq:su4_decomposition}
\begin{split}
    \text{SU}(4)_{i,j}(\params) =\  &\text{U}(\new{2})_i(\theta_{1:3})\times\text{U}(\new{2})_j(\theta_{4:6})\times\\
    &\text{XX}_{i,j}(\theta_7)\times\text{YY}_{i,j}(\theta_8)\times\text{ZZ}_{i,j}(\theta_9)\times\\
    &\text{U}(\new{2})_i(\theta_{10:12})\times\text{U}(\new{2})_j(\theta_{13:15}),
\end{split}
\end{align}
with a parameter vector $\params$ of length $15$. The notation $\theta_{l:l+2}$ for $l=1;4;9;13$, refers to the three elements in the $\params$ vector that go into each U$(\new{2})$ gate, e.g., $\theta_{1:3} = [\theta_1, \theta_2, \theta_3]$. We note that PQCs with only \sufour gates are redundantly parametrized because of the application of consecutive U$(\new{2})$ gates to the same qubit. When implementing on hardware, one may first compile the circuit into hardware-native gate sets and then combine trivially redundant gates.

In the case of the MPS initialized PQCs, the decomposed MPS tensors map to a U$(4)$ rotation, which includes a global phase $e^{-i\phi}$ with $\phi\in[0, 2\pi]$. The U$(4)$ were then decomposed into a trainable \sufour rotation with a fixed global phase via the KAK-decomposition:
\begin{equation}
    \text{U}(4) \xrightarrow{\text{KAK}} \text{SU}(4)(\params)\times e^{-i\phi}
\end{equation}

The \sufour gates in the MPS initialized simulations that were not part of the linear layers, i.e., the additional gates that were used to extend the PQC layer to an all-to-all topology, were initialized by sampling \new{the parameters from a} normal distribution $\mathcal{N}$ with zero mean $\mu=0$ and a small standard deviation $\sigma=0.01$. This was also done for all gates in the near-identity initializations in Fig.~\ref{fig:training_results}. The only exceptions were the additional gates in the VQE simulation \new{in Fig.~\ref{fig:training_results} and the large-scale gradient simulations in Fig.~\ref{fig:bas_gradient_scaling},} where $\sigma=0$ was chosen. \new{The former was done} because of the sensitivity of the energy loss in Eq.~\eqref{eq:vqe_energy} at these small scales\new{, whereas the latter was for computational feasibility of the gradient calculation.}

\subsection{PQC simulation}\label{apx:details_PQCsimulation}
All PQCs in this work were simulated using the Qulacs~\cite{qulacs2021} quantum circuit simulator through the interface provided by the \orquestra platform~\cite{Orquestra2022}. All results utilized exact statevector simulation of the quantum states $\psi_\params$ and the corresponding probabilities $q_\params(\textbf{x})$ described in Secs.~\ref{ssec:methods_QCBM} \&~\ref{ssec:methods_VQE}.

\subsection{PQC optimization}\label{apx:details_PQCoptimization}
The parameters of the PQC models in Sec.~\ref{sec:results} were optimized using a Python implementation of the CMA-ES~\cite{hansen1996adapting,hansen2019pycma} optimizer through the interface provided by the \orquestra platform~\cite{Orquestra2022}. In the case of the QCBM training, the initial step size was chosen as $\sigma_{\text{cma}}=10^{-2}$ in all depicted cases. For the VQE optimization, we found the energy error at the scales $10^{-1} - 10^{-2}$ to be very sensitive to changes in parametrization. Therefore, we chose $\sigma_{\text{cma}}=10^{-2}$ for the random and near-identity initializations, and $\sigma_{\text{cma}}=7.5\cdot 10^{-3}, \sigma_{\text{cma}}=5.0\cdot 10^{-3}$, and $\sigma_{\text{cma}}=2.5\cdot 10^{-3}$ for the MPS initialized modes with $\chi=2, \chi=4$, and $\chi=8$, respectively. The population sizes $\lambda_{\text{cma}}$ were always chosen to be $\lambda_{\text{cma}}=20$, meaning that each iteration in Fig.~\ref{fig:gradient_results} corresponds to $20$ quantum circuit simulations. In addition to a limit to the number of optimization steps, i.e., 10000, 15000, and 15000 for the PQCs in Fig.~\ref{fig:training_results} (for Cardinality, BAS, and the Heisenberg model respectively), we also set a loss tolerance of $5\cdot10^{-4}$ which may stop the optimization if differences of loss values between steps stay below this threshold. This can be observed in the $\chi=8$ example on the right tile of Fig.~\ref{fig:training_results}.

\subsection{Gradients}
The gradients with respect to the KL divergence loss (see Eq.~\eqref{eq:KL_divergence}) for Fig.~\ref{fig:gradient_results} were calculated using a finite-distance gradient estimator 
\begin{equation}
    \frac{\partial \mathcal{L}(\params)}{\partial \theta_l} = \frac{\mathcal{L}(\params+\epsilon\cdot\theta_l) - \mathcal{L}(\params-\epsilon\cdot\theta_l)}{2\epsilon}
\end{equation}
with $\epsilon=10^{-8}$ and for parameter index $l=8$ in  Eq.~\eqref{eq:su4_decomposition}.

\section{MPS decomposition}\label{apx:MPS_decomposition}
Our MPS decomposition protocol used throughout this work was proposed and demonstrated in Ref.~\cite{rudolph2022MPSdecomposition}, and combines the analytical decomposition technique in Ref.~\cite{ran2020encoding} with intertwined optimization steps of the created unitaries using TN techniques on classical computers. Concretely, for a target MPS wavefunction $|\psi_{\chi_\text{max}}\rangle$, the goal of the decomposition protocol is to output a sequence of circuit layers, i.e., $\prod_{i=k}^1 U^{(i)}$, where $U^{(i)}$ represent the two-qubit unitaries in the linear quantum circuit layer $i$, such that the fidelity 
\begin{equation}\label{eq:decomposition_fidelity}
\begin{aligned}
    f\big(\{U^{(i)}\}_i\big) & = \big | \langle 0^{\otimes N}| \prod_{i=k}^1 U^{(i)\dagger} |\psi_{\chi_\text{max}}\rangle \big |\\
    & = \big | \langle 0^{\otimes N}| U^{(k)\dagger}\cdots U^{(1)\dagger} |\psi_{\chi_\text{max}}\rangle \big |
\end{aligned}
\end{equation}
is maximized. The number of layers $k\leq K$ is chosen to be less or equal than a predetermined circuit depth limit $K$, or until a target fidelity $\hat{f}$ is achieved via $f\big(\{U^{(i)}\big)\}\geq\hat{f}$. The layer indexing convention from $k$ to $1$ is intentional as the decomposition protocol creates circuit layers ``from the MPS backwards'', such that the newest layer is implemented first in the quantum circuit.

The decomposition protocol sequentially creates circuit layers $U^{(i)}$ via the truncation and disentangling technique described in Ref.~\cite{ran2020encoding}. All existing circuit layers $\prod_{i=k^{'}}^1 U^{(i)}$ for a given decomposition iteration $k^{'} < k$ are optimized via the constrained optimization algorithm described in Ref.~\cite{shirakawa2021automatic}. That algorithm leverages the \textit{singular value decomposition} (SVD) and the fact that, for any matrix $\mathcal{M}$ that is decomposed via SVD, i.e., $\mathcal{M} = \mathcal{U}\cdot\mathcal{S}\cdot\mathcal{V}$, the product $\mathcal{U}\cdot\mathcal{V}$ represents the unitary matrix that best approximates $\mathcal{M}$. Ref.~\cite{rudolph2022MPSdecomposition} shows that this protocol is very effective at decomposing MPS into a low number $k$ of unitary circuit layers, especially with a limited computational budget for the optimization steps.
The precision of the decomposition, i.e., the fidelity $f\big(\{U^{(i)}\}_i\big)$ in Eq.~\eqref{eq:decomposition_fidelity}, can be sequentially improved with additional circuit layers and optimization steps. Additionally, unlike the method proposed in Ref.~\cite{dborin2022pretraining}, the protocol is not hindered by cumulative approximation error build-up when only a limited number of circuit layers are employed.

\clearpage

\end{document}